# Assessing workflow impact and clinical utility of AI-assisted brain aneurysm detection: a multi-reader study

Tommaso Di Noto[a*], Sofyan Jankowski[a], Francesco Puccinelli[a], Guillaume Marie[a], Sebastien Tourbier[a], Yasser Alemán-Gómez[a], Oscar Esteban[a], Ricardo Corredor-Jerez[b], Guillaume Saliou[a], Patric Hagmann[a], Meritxell Bach Cuadra[ac], Jonas Richiardi[a]

a. Department of Radiology, Lausanne University Hospital (CHUV) and University of Lausanne (Unil), Rue du Bugnon 46, 1005 Lausanne, Switzerland
b. Siemens Healthineers International AG, Swiss Innovation Hub, EPFL Innovation Park, 1015, Lausanne, Switzerland
c. CIBM Center for Biomedical Imaging, EPFL, 1015, Lausanne, Switzerland
* Corresponding author (tommydino@hotmail.it)

## Abstract

Despite the plethora of AI-based algorithms developed for anomaly detection in radiology, subsequent integration into clinical setting is rarely evaluated. In this work, we assess the applicability and utility of an AI-based model for brain aneurysm detection comparing the performance of two readers with different levels of experience (2 and 13 years). We aim to answer the following questions: 1) Do the readers improve their performance when assisted by the AI algorithm? 2) How much does the AI algorithm impact routine clinical workflow? We reuse and enlarge our open-access, Time-Of-Flight Magnetic Resonance Angiography dataset (N=460). We use 360 subjects for training/validating our algorithm and 100 as unseen test set for the reading session. Even though our model reaches state-of-the-art results on the test set (sensitivity=74%, false positive rate=1.6), we show that neither the junior nor the senior reader significantly increase their sensitivity (p=0.59, p=1, respectively). In addition, we find that reading time for both readers is significantly higher in the "AI-assisted" setting than in the "Unassisted" (+15 seconds, on average; $p = 3 \times 10^{-4}$ junior, $p = 3 \times 10^{-5}$ senior). The confidence reported by the readers is unchanged across the two settings, indicating that the AI assistance does not influence the certainty of the diagnosis. Our findings highlight the importance of clinical validation of AI algorithms in a clinical setting involving radiologists. This study should serve as a reminder to the community to always examine the real-word effectiveness and workflow impact of proposed algorithms.

# 1 Introduction

### 1.1 Clinical background

Unruptured intracranial aneurysms (UIAs) are abnormal dilatations of cerebral arteries [1]. The prevalence of UIAs in the adult population ranges between 1% and 5% [1], [2], and their rupture is the predominant cause of nontraumatic subarachnoid hemorrhages (SAH) [3]. The mortality rate of SAH is around 40% and only half of post-SAH patients return to independent life [4]. Time-of-flight magnetic resonance angiography (TOF-MRA) is a non-invasive imaging technique that has found widespread clinical application to locate UIAs [5]. If spotted early, UIAs can be monitored over time and considered for treatment if their risk of rupture becomes too high [6]. Radiologists' visual detection of UIAs based on TOF-MRA scans presents several limitations. First, sensitivity for detecting small UIAs (<5mm) can be as low as 35% [7], especially for inexperienced radiologists. Second, about 10% of all UIAs are missed during routine clinical examination [8]. Third, even medium-sized UIAs are hardly detected on maximum intensity projection (MIP) images (technique that enhances vascular structures by projecting the highest intensity voxels along a viewing direction) because of overlap with adjacent vessels and unusual locations in the vasculature [9]. Last, factors like fatigue and satisfaction of search [10] can further increase the number of missed UIAs. For these reasons, the development of a computer-aided detection (CAD) tool [11] that supports clinicians in detecting UIAs would be highly beneficial, especially considering the increasing workload of radiologists in their daily clinical routine [12], [13].

### 1.2 Related works

Several studies have explored different automated methods for detecting UIAs in TOF-MRA. In particular, starting from around 2016, the vast majority of studies adopted DL algorithms, which permitted to achieve unprecedented performances and became the standard for UIA detection [14], [15], [16], [17], [18], [19]. A significant leap forward for the community was brought by the Aneurysm Detection And segMentation Challenge (ADAM) [20]. This allowed to obtain a fair comparison across methods, and it revealed the true difficulty of the detection task, considering that none of the top-5 algorithms exceeded a sensitivity of 70%.

Among the research groups cited above, only a few assessed, in follow-up studies ([21], [22], [23], [24], [25]), the utility and applicability of their CAD tool in a clinical setting. However, Faron et al. [25] only presented patients with UIAs to the human readers (i.e., no healthy controls), a scenario which does not reflect clinical practice. In [21], [22], the authors included control subjects, but only selected patients with one UIA (i.e., no patients with multiple UIAs). Moreover, they informed the readers about this selection which might have led to an expectation bias [26]. The work by Štepán-Buksakowska et al. [23] is limited by the low sample size (only 9 patients with UIAs included) and excessively high prediction rate (7.2 hits per subject). Last, the work by Miki et al. [24] slightly differs from the others because, rather than investigating the performance of human readers with and without the CAD system on repeated subjects (within-subjects analysis), the authors evaluated detection performances of the readers only with the CAD.

Differently from these studies, in this work we conducted a within-subject analysis to assess the utility of our CAD aneurysm detection algorithm (described in [14]) in a controlled clinical setting, including both patients with one or more UIAs and control subjects. Specifically, the questions that we aimed to answer are:

1. How does the CAD affect the sensitivity and the specificity of the readers?
2. How does the CAD impact the workflow of the readers in terms of time and confidence?

# 2 Materials & Methods

## 2.1 In-house Dataset

This study was approved by the regional ethics committee; written informed consent was waived.

**Training dataset -** We retrieved a retrospective cohort of **N=279** subjects (151 with one or more UIAs, 128 control subjects without UIAs) who underwent TOF-MRA scans at the Lausanne University Hospital between 2010 and 2015. All these training subjects were also used in [14], where we investigated the top-performing DL configuration for UIA detection. The subjects were included in a consecutive manner by looking at the corresponding radiology reports. Indications for the included subjects were, among others, meningitis, suspected stroke, tinnitus, aphasia, vestibular dysfunctions, persistent hiccup, post-stroke follow-up, migraine, sudden vision loss, facial paresis and aneurysm follow-up. We excluded patients with ruptured/treated aneurysms, totally thrombosed aneurysms, or infundibula (dilatations of the origin of an artery).

**Validation dataset** – We retrieved **N=81** TOF-MRA subjects (65 controls without UIAs and 16 patients with UIAs) scanned in 2015 at the Lausanne University Hospital. Inclusion and exclusion criteria were identical to those of the training dataset. These 81 subjects were used to find the best transfer learning configuration to maximize detection performances (details in section 2.3).

**Test dataset -** We retrieved **N=100** additional TOF-MRA subjects (63 controls without UIAs and 37 patients with UIAs) scanned in 2015 at the Lausanne University Hospital. These are separate subjects and do not overlap neither with the training nor with the validation dataset. Inclusion and exclusion criteria were identical to those of the training dataset. These 100 subjects were solely used to assess the clinical utility of our CAD tool in a within-subject reading (details in section 2.4).

Table 1 summarizes the main demographic information for our datasets. MRI acquisition parameters are reported as Supplementary Materials (Table A). The datasets were depersonalized and organized according to the Brain Imaging Data Structure (BIDS) specification [27]. The original portion of the training dataset (i.e., all subjects used in [14]) is available on OpenNeuro [28] at https://openneuro.org/datasets/ds003949.

**Table 1**. Demographics for training, validation and test datasets. UIA rate = # aneurysms / # subjects. Patients = subjects with UIAs. Controls = subjects without UIAs. Age presented as mean ± standard deviation. N=number of subjects; M=males; F=females. Mann-Whitney U test to compare age between patients and controls. Chi-squared test to compare sex counts between patients and controls.

| | | Patients | Controls | Test, *p* value | Whole Sample |
|---|---|---|---|---|---|
| **Training dataset** | **N** | 151 | 128 | / | 279 |
| | **Age (y)** | 56±14 | 46±18 | $U = 65090\ p = 2 \times 10^{-6}$ | 52±17 |
| | **Sex** | 51M, 100F | 61M, 67F | $\chi^2 = 4.9, p = 0.02$ | 112M, 167F |
| | **UIA rate** | 1.16 | 0 | / | 0.63 |
| **Validation dataset** | **N** | 16 | 65 | / | 81 |
| | **Age (y)** | 55±12 | 53±16 | $U = 520\ p = 1{,}0$ | 54±16 |
| | **Sex** | 6M, 10F | 31M, 34F | $\chi^2 = 0.2, p = 0.6$ | 38M, 43F |
| | **UIA rate** | 1.68 | 0 | / | 0.33 |
| **Test dataset** | **N** | 37 | 63 | / | 100 |
| | **Age (y)** | 58±11 | 53±15 | $U = 930\ p = 0{,}09$ | 55±14 |
| | **Sex** | 17M, 20F | 38M, 25F | $\chi^2 = 1.4, p = 0.2$ | 55M, 45F |
| | **UIA rate** | 1.27 | 0 | / | 0.47 |

### 2.2 ADAM dataset

To find the optimal transfer learning configuration for our model (details in section 2.3), we also used the ADAM dataset, composed of 113 subjects (20 controls and 93 patients with 125 UIAs).

### 2.3 Aneurysm Characteristics and Manual Annotations

The 151 UIA patients in our **training dataset** presented a total of 176 UIAs. These 176 UIAs were manually annotated by one radiologist (author G.M.) with 2 years of experience in neuroimaging (details in [14]). 161/176 (92%) of the UIAs in our training dataset are saccular, while 15/176 (8%) are fusiform.

The 16 UIA patients in our **validation dataset** presented 27 saccular UIAs (100%). These were manually annotated with ITK-SNAP (v.3.6.0) [29] by a second radiologist (author P.H.) with over 18 years of experience in neuroimaging. For these 27 UIAs, the radiologist drew the voxel-wise aneurysm labels scrolling through the axial planes of the TOF-MRA scans.

The 37 UIA patients in our **test dataset** presented 47 saccular UIAs (100%). All these 47 UIAs were manually annotated once again by P.H. with ITK-SNAP.

All TOF-MRA exams (training, validation, test - both patients and controls) were double-checked by a third senior neuroradiologist (author G.S.) with over 15 years of experience in order to exclude any false positive detection or false negative (missed UIA).

Table 2 illustrates the locations and sizes of all the saccular UIAs in our training, validation and test dataset, grouped according to the PHASES score [6]. This is a clinical score used to assess the 5-year risk of rupture of saccular UIAs. Since not all PHASES criteria were available, we only recorded aneurysm size, location and patient age as in [14]. The aneurysm locations in the PHASES score are: 1) ICA (Internal Carotid Artery), 2) MCA (Middle Cerebral Artery), and 3) ACA (Anterior Cerebral Arteries)/Pcom (Posterior communicating artery)/Posterior. Fusiform UIAs and extracranial carotid artery UIAs were excluded from the table (but not from the whole reading analysis) just because they do not meet the inclusion criteria of the PHASES score.

### 2.4 Data Processing, Network & Training

In this section, we briefly summarize the preprocessing pipeline and the DL architecture that were developed in [14] and that we adopted in this study.

**Preprocessing -** Several preprocessing steps were carried out for each subject. First, we performed skull-stripping with FSL (v. 6.0.1) [30]. Second, we applied N4 bias field correction with SimpleITK (v. 1.2.0) [31]. Third, we resampled all volumes to a median voxel spacing (0.39x0.39x0.55 mm). Last, a probabilistic vessel atlas built from multi-center MRA datasets [32] was co-registered to each patient's TOF-MRA using ANTS (v. 2.3.1) [33] (details in Supplementary Materials of [14]). In this work, we only used the vessel atlas to carry out the anatomically-informed sliding window at inference time since this configuration led to higher performances (see Table 5 from [14]).

**Network & Training –** The DL model used in this study is a custom 3D UNET, inspired by the original work [34], with upsample layers in the decoding branch rather than transpose convolutions. We used 3D TOF-MRA patches as input to our network. We set the size of the input patches to 64x64x64 voxels. The output of the network is, for every input patch, a corresponding probabilistic patch where non-zero voxels correspond to potential aneurysm candidates. Further details about the architecture can be found in [14].

Differently from [14], here we explored distinct transfer learning (TL) configurations during training. In general, TL is the branch of machine learning where knowledge acquired from a specific task or domain (source) is exploited to solve a downstream, related task (target) [35]. In this work, we considered as source domain the ADAM dataset, which was used for pretraining, and as target domain the training dataset described in section 2.1. We explored 4 different TL configurations: mixed training, encoder finetuning, decoder finetuning, and finetuning of all layers. Details about these experiments are provided in Section A of the Supplementary Materials. After training with these four TL approaches, we ran inference on the validation subjects and picked the configuration that showed highest detection performance. Training and inference were performed with Tensorflow 2.4.0 and a GeForce RTX 2080TI GPU with 11 GB of SDRAM. We make the code used for this manuscript available at *https://github.com/connectomicslab/AI-Assisted-Aneurysm-Detection*.

**Table 2**. Locations and sizes of the saccular UIAs divided according to the PHASES score criteria for the training, validation and test datasets. ICA = Internal Carotid Artery, MCA = Middle Cerebral Artery, ACA = Anterior Cerebral Arteries, Pcom = Posterior communicating artery, Posterior = posterior circulation. $d$ = maximum diameter. **N.B.** the total number of UIAs slightly differ from those in the text because the table neglects fusiform aneurysms which cannot be considered for the PHASES score.

| | | | **Count** | **%** |
|---|---|---|---|---|
| **Training Dataset** | **Location** | ICA | 30 | 20.2 (30/148) |
| | | MCA | 49 | 33.1 (49/148) |
| | | ACA/Pcom/Posterior | 69 | 46.7 (69/148) |
| | **Size** | $d \leq 7\ mm$ | 141 | 95.3 (141/148) |
| | | $7-9{,}9\ mm$ | 2 | 1.3 (2/148) |
| | | $10-19{,}9\ mm$ | 5 | 3.4 (5/148) |
| | | $d \geq 20\ mm$ | 0 | 0 (0/148) |
| **Validation Dataset** | **Location** | ICA | 12 | 44.4 (12/27) |
| | | MCA | 8 | 29.6 (8/27) |
| | | ACA/Pcom/Posterior | 7 | 26.0 (7/27) |
| | **Size** | $d \leq 7\ mm$ | 26 | 96.3 (26/27) |
| | | $7-9{,}9\ mm$ | 0 | 0 (0/27) |
| | | $10-19{,}9\ mm$ | 1 | 3.7 (1/27) |
| | | $d \geq 20\ mm$ | 0 | 0 (0/27) |
| **Test Dataset** | **Location** | ICA | 20 | 44.4 (20/45) |
| | | MCA | 14 | 31.1 (14/45) |
| | | ACA/Pcom/Posterior | 11 | 24.5 (13/45) |
| | **Size** | $d \leq 7\ mm$ | 43 | 95.5 (43/45) |
| | | $7-9{,}9\ mm$ | 2 | 4.5 (2/45) |
| | | $10-19{,}9\ mm$ | 0 | 0 (0/45) |
| | | $d \geq 20\ mm$ | 0 | 0 (0/45) |

### 2.5 Experimental Setup

Two radiologists (hereafter referred to as readers) volunteered to undergo the task of UIAs detection in TOF-MRA scans. One reader (F.P., hereafter referred to as *senior*) has 13 years of experience in neuroimaging, while the second reader (S.J., hereafter referred to as *junior*) has 2 years of experience in neuroimaging. These two readers are distinct from those who created the ground truth annotations, but work in the same department. We set up the study as a paired, within-subject image reading task: the two readers were asked to visually inspect the 100 test TOF-MRA volumes under two different settings:

1 **Unassisted**: the readers had access only to the original TOF-MRA image and they could freely explore all three views (axial, sagittal, coronal), as well as the MIP reconstruction.

2 **AI-assisted**: the readers had access both to the original TOF-MRA image (3 views + MIP) and to the same image that contained potential aneurysm candidates generated by our top-performing DL model. We denote this image as *overlay*. Both images (original and overlay) were available to the readers and could be viewed side by side. The readers could freely decide which image to examine first. Also, they could decide whether to consider or neglect the aneurysm(s) candidates shown in the overlay. The overlay image was generated with MeVisLab (v.3.4.2) [36]: each UIA candidate was shown as a red contour and with an associated, uncalibrated probability (details in Section B of the Supplementary Materials). Figure 1 depicts the AI-assisted setting for two test patients.

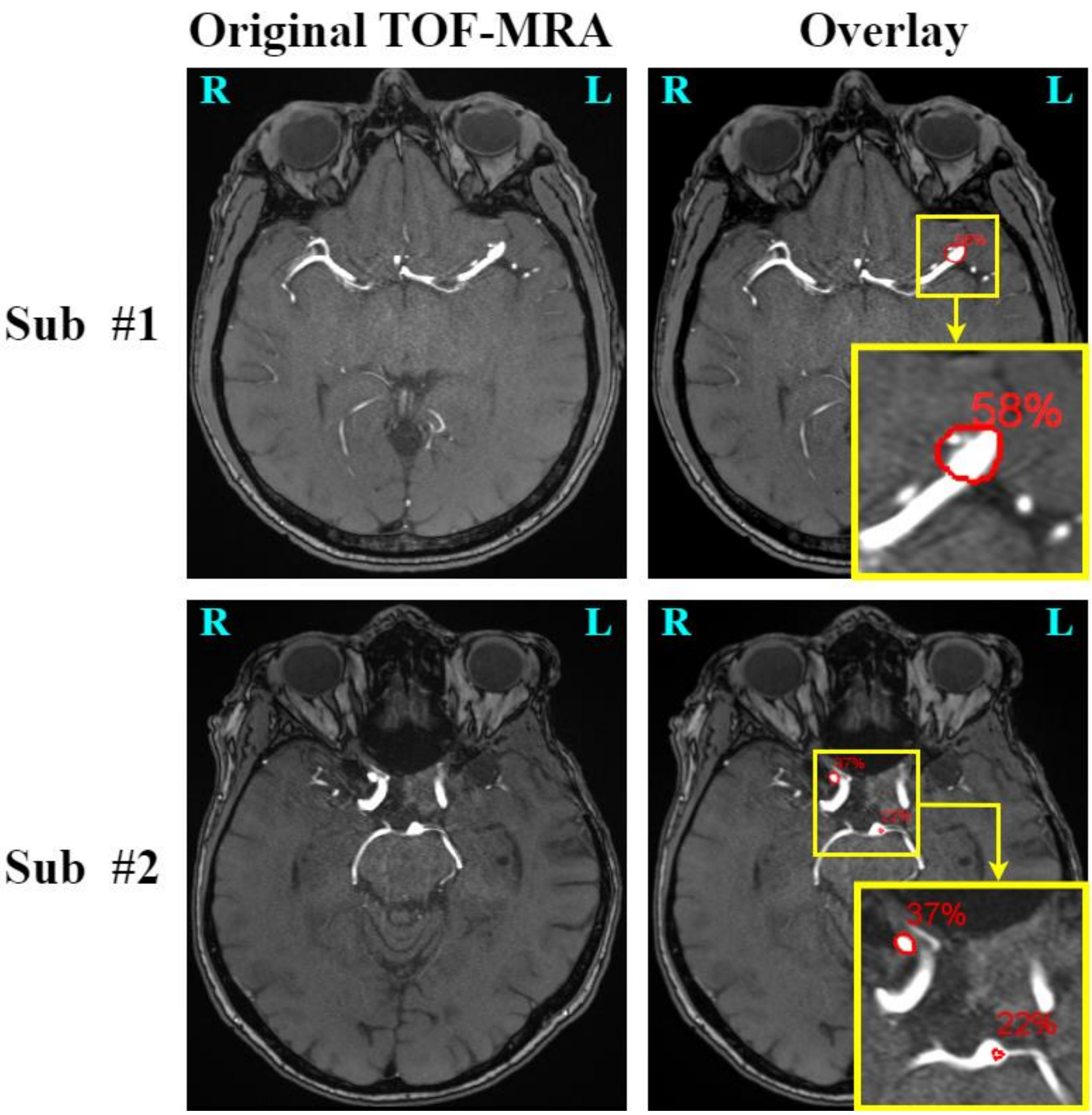

***Figure 1***. *Two examples of the AI-assisted setting.* ***First row****: axial view of a 53-year-old male patient with an aneurysm located in the M2 segment of the middle cerebral artery. The aneurysm is correctly detected in the overlay image (second column).* ***Second row****: axial view of a 70-year-old female patient with a correctly detected aneurysm in the right carotid siphon (37%) and a false positive prediction in the tip of the basilar artery (22%). Note: the yellow box is only used for the zoom in the Figure, but was not present in the overlay.*

We conducted two reading sessions: in the first session, the 100 test patients were randomly assigned either to the Unassisted or to the AI-assisted setting, while in the second session, the opposite scenario was presented to the readers. For instance, a patient that was inspected with the Unassisted setting on the first reading session was then assessed under the AI-assisted setting on the second session, and vice-versa. Since patient order might influence detection performance, for instance due to fatigue or boredom, we kept the same random order for both readers. The time interval between the two reading settings was approximately 40 days for both the junior and the senior radiologist, depending on their availability. Due to time constraints, multiple sessions were needed to complete the settings (unassisted, assisted), both for the senior and for the junior. The readers were blinded to the performance levels of the CAD system, and to the exact prevalence of UIAs in the test set, to avoid any expectation bias [26]. The following instructions were provided to each reader:

"*You will be presented 100 TOF-MRA scans. Some belong to healthy controls and some to patients with unruptured intracranial aneurysm(s). For each scan, please indicate the presence and location of aneurysms, if any was spotted. For the locations, refer to the list provided to you. For every aneurysm detected, also indicate a confidence score using a scale from 1 (very low confidence of being an aneurysm) to 10 (very high confidence: unequivocal aneurysm). There is no time limit per scan, but try to simulate reading times of a routine clinical day in your department.*"

The complete list of locations that was given to the readers is provided as Supplementary Material (Table B). The reading sessions were carried out with the Carestream PACS system (v.12.2.1), which is the DICOM viewer routinely used at Lausanne University Hospital. Before the actual reading on the 100 test subjects, each reader underwent a training session with 2 cases belonging to the validation set to familiarize themselves with the interface of the AI-assisted setting.

### 2.5 Statistical analysis

To evaluate the effectiveness of the AI-based CAD system, we ran 4 tests with hypotheses specified a priori:

**Test 1:** we hypothesized that the sensitivity of the junior radiologist under the AI-assisted setting would be significantly higher than his sensitivity under the Unassisted setting. In other words, we expected the junior to miss fewer aneurysms under the AI-assisted setting. This test was run only for the 37/100 test patients that have one or more UIAs. Since the majority of patients (~80%) has only one UIA [37], we considered sensitivity aneurysm-wise, and not patient-wise. For patients with multiple UIAs, we considered each aneurysm as independent. With this simplification, the sensitivity analysis becomes binary (aneurysm found vs. aneurysm not found) and testing can thus be performed through a McNemar's test [38]. This test is suitable for paired (i.e., within-subject) settings like ours (Unassisted vs. AI-assisted). The contingency matrix is defined as:

| | | **AI-assisted** | |
|---|---|---|---|
| | | True Positives (TPs) | False Negatives (FNs) |
| **Unassisted** | True Positives (TPs) | a | b |
| | False Negatives (FNs) | c | d |

where cell "a" contains the number of UIAs found by the reader under both settings, "b" contains UIAs found by the reader only under the Unassisted setting, "c" contains UIAs found only under the AI-assisted setting, and "d" contains UIAs that the reader missed under both settings. From such matrix, we ran the McNemar's test with the R function **mcnemar.test**, setting continuity correction to false and a significance threshold $\alpha = 0.05$.

**Test 2:** we hypothesized that the sensitivity of the senior radiologist under the AI-assisted setting would be statistically equivalent to his sensitivity under the Unassisted setting. For this test, we used the same assumptions and analysis of **Test 1**.

**Test 3:** we hypothesized that the specificity of the junior radiologist would be statistically equivalent between the two settings (Unassisted and AI-assisted). In other words, we expected the junior to be able to discard false positive predictions provided by the DL model. This test was run only for the 63/100 test controls that do not have UIAs, as similarly performed in [16]. The specificity analysis was run patient-wise, in order to make the problem binary. A control subject is considered a true negative if no UIAs are predicted by the reader, while it is considered a false positive if one or more UIAs are predicted. As for **Test 1** and **Test 2**, we compared the two settings (Unassisted vs. AI-Assisted) with a McNemar's test. This contingency matrix has the same form as above, but counts now reflect patients instead of UIAs. As for **Test 1** and **Test 2**, we ran McNemar's test with R.

**Test 4:** we hypothesized that the specificity of the senior would be statistically comparable between the two settings (Unassisted and AI-assisted). For this test, we used the same assumptions and analysis of **Test 3**.

**Test 5:** we compared the sensitivity (only for patients with UIAs) and the specificity (only for patients without UIAs) of the two readers under the same setting (i.e. Unassisted junior vs. Unassisted senior, and AI-assisted junior vs. AI-assisted senior), using two-tailed Wilcoxon signed-rank tests.

## 2.6 Reading Timing and Confidence

The readers were timed by an external observer (T.D.N.) in order to streamline the reading sessions and avoid extra work for the readers. The stopwatch was started when the reader had the TOF image (for the Unassisted setting) and the TOF + overlay (for the AI-assisted) open in front of him. Then, the stopwatch was stopped as soon as the reader closed the case and was ready to report all the UIAs found, or the lack thereof for subjects deemed as controls.

In addition, as mentioned in the instructions above, the readers were asked, for every spotted aneurysm, to report a confidence score from 1 to 10 (1: very low confidence, 10: very high confidence).

To assess the changes in reading time under the various reading settings (junior vs. senior, Unassisted vs. AI-assisted), and to compare the confidence scores reported by the readers, we performed multiple two-sided Wilcoxon signed-rank tests. These tests were run with SciPy (v.1.12), setting a significance threshold $\alpha = 0.05$.

# 3 Results

### 3.1 Sensitivity analysis: DL model and readers

The DL model that achieved the highest detection performances on the validation set was the one in which we finetuned all layers on the validation dataset, after pretraining on the source ADAM data. When applied to the unseen test set, this model achieved a sensitivity of 74% (35/47 UIAs detected). For perspective, this is on par with the top algorithm in the ADAM challenge (see live leaderboard; the website is currently down but results can be seen with the *waybackmachine* to June 26, 2022). Nonetheless, the data are different and thus performance is difficult to compare.

The contingency matrices of the sensitivity analysis for the two readers are shown in Figure 2. Under the AI-assisted setting, the junior radiologist increased his sensitivity to 83% compared to a sensitivity of 78% in the Unassisted setting. Conversely, the senior radiologist attained a sensitivity of 87% regardless of the reading setting. When running the McNemar for **Tests 1** and **2** presented in section 2.5 (i.e., sensitivity analysis – AI-assisted vs. Unassisted), we found no significant difference neither for the junior ($\chi^2 = 0.2,\ p = 0.59$), nor for the senior reader ($\chi^2 = 0,\ p = 1$).

| Junior | | AI-Assisted: True Positives (TP) | AI-Assisted: False Negatives (FN) |
|---|---|---|---|
| Unassisted | True Positives (TP) | 31 | 6 |
| | False Negatives (FN) | 8 | 2 |

**Sensitivity**
AI-assisted = 83% (39/47 UIAs found)
Unassisted = 78% (37/47 UIAs found)

| Senior | | AI-Assisted: True Positives (TP) | AI-Assisted: False Negatives (FN) |
|---|---|---|---|
| Unassisted | True Positives (TP) | 37 | 4 |
| | False Negatives (FN) | 4 | 2 |

**Sensitivity**
AI-assisted = 87% (41/47 UIAs found)
Unassisted = 87% (41/47 UIAs found)

***Figure 2***. *Contingency matrices of sensitivity analysis (AI-assisted vs. Unassisted). Left: junior radiologist. Right: senior radiologist. UIAs: Unruptured Intracranial Aneurysm(s). TP: true positive UIAs. FN: false negative (i.e., missed UIAs).*

**Junior radiologist error analysis** – Out of the 8 UIAs that were missed by the junior radiologist during the AI-assisted setting, 4 were located in the Middle Cerebral Artery (MCA) and 4 in the Internal Carotid Artery (ICA). All of them were smaller than 3.5 mm, except for a large, 7-mm wide aneurysm. In the Unassisted setting, the junior missed 10 UIAs: 4 MCA, 3 ICA, and 3 in the ACA/Pcom/Posterior regions. All these missed UIAs were smaller than 3.1 mm. We also noticed a satisfaction-of-search effect (i.e., detecting only 1 UIAs for patients with multiple UIAs): specifically, 25% (2/8) of the UIAs missed during the AI-assisted setting belonged to patients with multiple UIAs, compared to an even higher 60% (6/10) in the Unassisted setting.

**Senior radiologist error analysis** – In the AI-assisted setting, the senior missed 4 ICA, 1 MCA, and 1 ACA/Pcom/Posterior aneurysms. These UIAs were all smaller than 2.4 mm, except for one 4 mm-wide UIA. In the Unassisted setting, the senior missed 3 ICA, 1 MCA, and 2 ACA/Pcom/Posterior aneurysms, all smaller than 3.1 mm. The satisfaction-of-search effect was even stronger for the senior with a 66% (4/6) ratio under the AI-assisted setting and a 50% (3/6) ratio in the Unassisted setting.

### 3.2 Specificity analysis: DL model and readers

Our top-performing DL model had an average false positive (FP) rate of 1.62 per subject on the unseen test set (N=100). For perspective, this is slightly higher than the first (0.40) and third (0.13) top algorithm of the ADAM challenge (live leaderboard), but much lower than the second team (FP rate = 4.03). Note that rankings reflected a combination of sensitivity and false positives and, once again, test datasets are different, thus comparisons should be interpreted with caution.

Figure 3 illustrates the contingency matrices of the specificity analysis for the two readers. The junior radiologist had a slightly lower specificity under the AI-assisted setting (68%) with respect to the Unassisted setting (71%). Similarly, the senior radiologist showed a specificity of 98% for the AI-assisted setting, and 100% for the Unassisted setting. When running the McNemar for **Tests 3** and **4** (i.e., specificity analysis – AI-assisted vs. Unassisted), we found no significant difference in specificity between the two settings, neither for the junior ($\chi^2 = 0.2,\ p = 0.65$) nor for the senior ($\chi^2 = 1,\ p = 0.3$).

**Junior**

| Unassisted \ AI-Assisted | True Negatives (TN) | False Positives (FP) |
|---|---|---|
| True Negatives (TN) | 34 | 11 |
| False Positives (FP) | 9 | 9 |

**Specificity**
AI-assisted = 68% (43/63 real TN)
Unassisted = 71% (45/63 real TN)

**Senior**

| Unassisted \ AI-Assisted | True Negatives (TN) | False Positives (FP) |
|---|---|---|
| True Negatives (TN) | 62 | 1 |
| False Positives (FP) | 0 | 0 |

**Specificity**
AI-assisted = 98% (62/63 real TN)
Unassisted = 100% (63/63 real TN)

***Figure 3****. Contingency matrices of specificity analysis (AI-assisted vs. Unassisted). Left: junior radiologist. Right: senior radiologist. TN: truly negative patients without UIAs. FP: control subjects for which at least one false positive UIA was predicted.*

**Junior radiologist error analysis –** Out of the 20 FP predictions reported by the Junior under the AI-Assisted setting, 40% (8/20) were located in the Internal Carotid Artery (ICA), 35% (7/20) in the ACA/Pcom/Posterior regions and 25% (5/20) in the Middle Cerebral Artery (MCA). With a similar trend, under the Unassisted setting, 55% (10/18) of the FPs were located in the ICA, 38% (7/18) in the ACA/Pcom/Posterior, and 5% (1/18) in the MCA.
**Senior radiologist error analysis –** The only FP reported by the Senior was under the AI-Assisted setting and was located in the Anterior Cerebral Artery.

### 3.3 Junior vs. Senior under the same setting

**Sensitivity analysis** - Although the senior radiologist consistently outperformed the junior (as shown in section 3.1), the differences were not significant, neither for the Unassisted (W=19, $p = 0.40$), nor for the AI-Assisted setting (W=24, $p = 0.25$).

**Specificity analysis** – Conversely, the differences in specificity were significantly different between the two readers with the senior once again consistently outperforming the junior, both for the Unassisted (W=0, $p = 2 \times 10^{-5}$), and the AI-assisted setting (W=0, $p = 1 \times 10^{-5}$).

### 3.4 Reading Timing and Confidence

#### 3.4.1 Reading Timing

The differences in reading time between the two readers and across the two settings are displayed in Figure 4. The median reading time of the junior reader under the AI-assisted setting (122 s) was significantly higher (W=1444, $p = 3 \times 10^{-4}$) than his reading time under the Unassisted setting (107 s). Similarly, the median reading time of the senior under the AI-assisted setting (103 s) was significantly higher (W=1255, $p = 3 \times 10^{-5}$) than his reading time under the Unassisted setting (89 s). When comparing the timing of the two readers within the same setting, the senior was significantly faster than the junior both for the AI-assisted (W=1575, $p = 0.001$) and for the Unassisted setting (W=1661, $p = 0.003$).

### 3.4.2 Confidence scores – Assisted vs. Unassisted

In Figure 5, we show the confidence scores reported by the two readers under both settings. To make the comparison more robust, here we consider all FP predictions including those belonging to patients with UIAs and not only those belonging to control subjects as done in section 3.2. When comparing all confidence scores (TP and FP combined) of the junior reader between the two settings (Assisted vs. Unassisted), we found no significant difference (W=409, $p = 0.43$), indicating that the junior reader's confidence was unaffected by AI assistance. Likewise, we observed a similar trend for the senior that also reported statistically comparable confidence scores across the two settings (W=40, $p = 0.42$).

### 3.4.3 Confidence scores – TPs vs. FPs

When comparing TPs and FPs of the junior within the same setting, we found that the confidence for TPs was significantly higher than the one for FPs, both for the AI-assisted (W=26, $p = 1 \times 10^{-5}$) and for the Unassisted setting (W=16.5, $p = 5 \times 10^{-6}$). This comparison was skipped for the senior since he only reported 1 FP across the two settings.

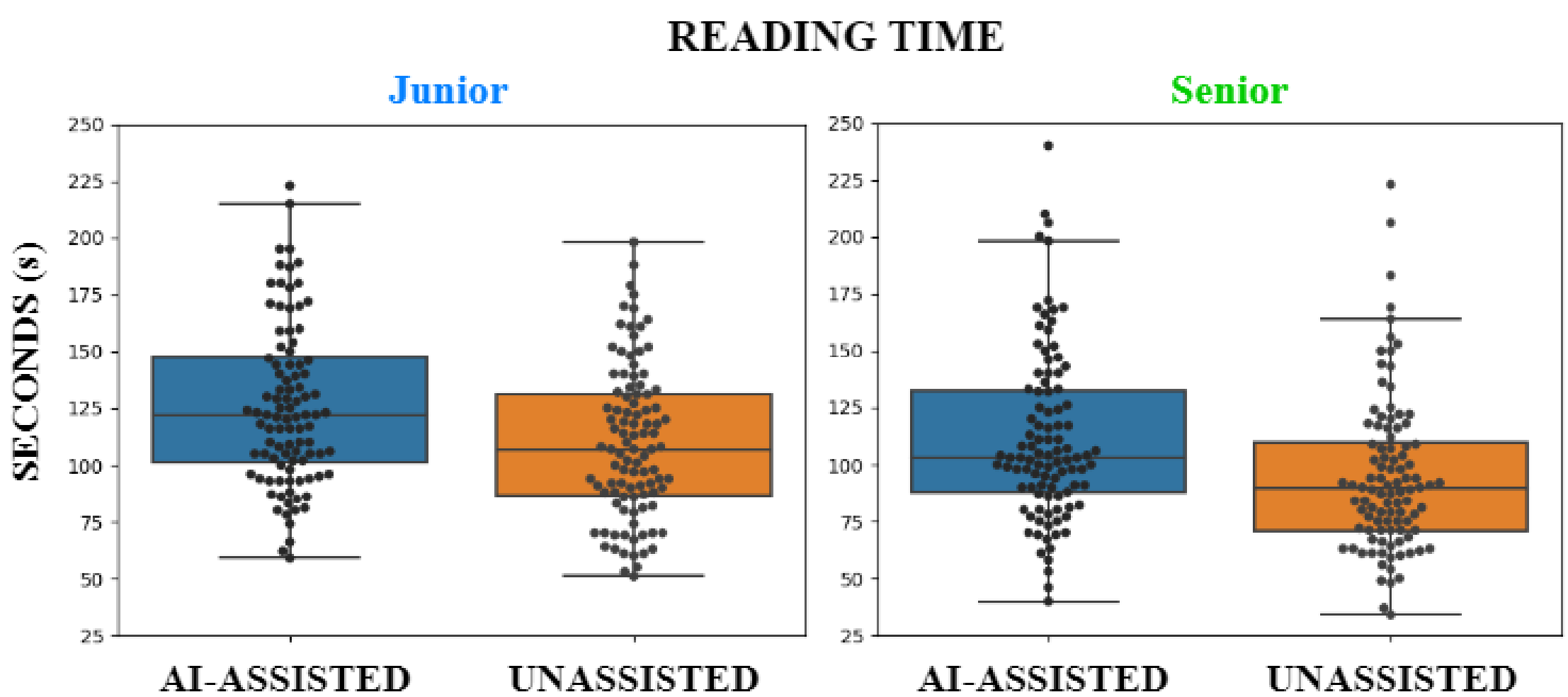

***Figure 4***. *Reading times across the two settings. Each dot represents the time spent by the reader to explore one subject. Left: junior radiologist. Right: senior radiologist.*

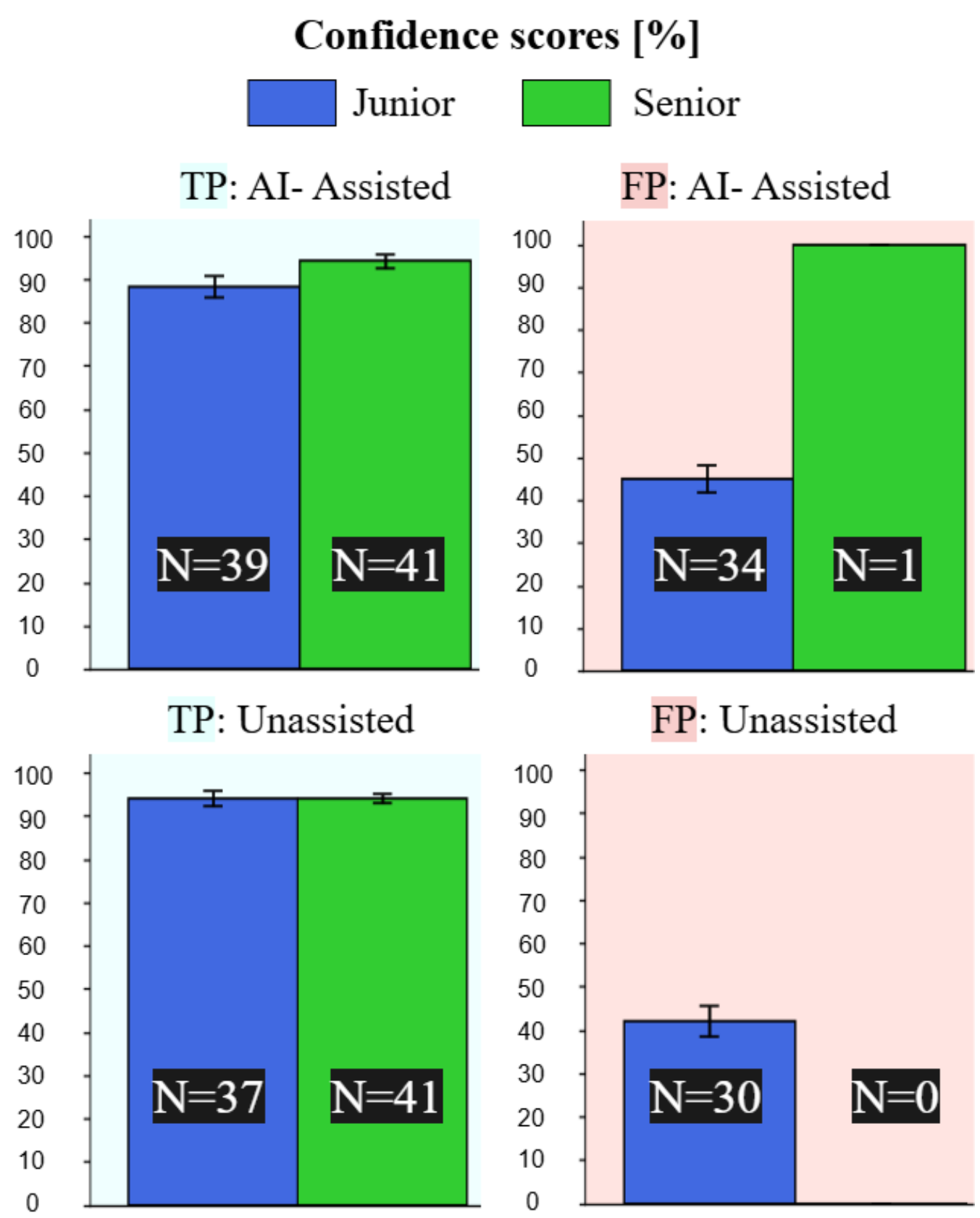

***Figure 5****. Confidence score reported by the readers under the two settings. The bar plots indicate the confidence in % and the error bars indicate the standard deviation.* ***Top left****: True Positives aneurysms for the AI-Assisted setting.* ***Top right****: False Positives predictions for the AI-assisted setting.* ***Bottom left****: True Positive aneurysms for the Unassisted setting.* ***Bottom right****: False Positives predictions for the Unassisted setting. Blue bars: Junior; green bars: Senior.*

# 4 Discussion

In this manuscript, we investigated the applicability of a CAD system for AI-assisted aneurysm detection on TOF-MRA images in a controlled clinical scenario. Specifically, we carried out a within-subject analysis and assessed diagnostic performances of two readers with different levels of experience with and without the support of the CAD tool.

Despite the multitude of academic papers being published in the field of DL in radiology, very few of the proposed algorithms undergo subsequent evaluation in clinical practice [39], [40]. In fact, it has been reported that this last step of model deployment still remains a major bottleneck for the integration of AI-based CAD systems in clinical routine [41]. Contrary to this trend, this work aimed at closing the algorithm life cycle by deploying our previously-developed CAD tool into a controlled clinical setting.

In the field of aneurysm detection, sensitivity is paramount since oversights can have dreadful consequences for patients. Even though our model showed state-of-the-art sensitivity on the unseen test set (i.e., 74%, is higher than ADAM challenge results with reasonable FP rate), this study demonstrated that such a performance is still much too low to assist the senior reader, given his reported sensitivity of 87% both under the Unassisted and AI-assisted settings. Similarly, despite the detection improvement observed for the junior reader (from 78% Unassisted to 85% AI-assisted), we found that this sensitivity gain was not significant. In light of these results, we posit that the sensitivity of any proposed CAD system for UIAs detection should be greater or equal than that attained by the senior. This level of sensitivity could be beneficial for inexperienced readers to increase their detection rate. Additionally, it could support senior readers acting as "second reader" during routine clinical practice.

Besides improving sensitivity, AI-based CAD tools in radiology should also improve the efficiency of existing workflows [42]. For this reason, the readers in this study were timed by an external observer to gain insights into changes in the reading process. As shown in section 3.4, reading times for both readers were statistically superior under the AI-assisted setting. We believe the explanation for this increase is twofold: on one hand, the mere fact of having to read an extra image in the AI-assisted setting (i.e., the *overlay*) increases reading time. On the other hand, the relatively high number of false positives predicted by the CAD tool (on average, 1.62 per case) also led to a substantial increase in reading time. While little can be done for the reading of the extra image (any CAD tool provides extra information for radiologists), we argue that future works targeting AI-assisted aneurysm detection should aim at reducing the FP rate, for instance using a dedicated network (subsequent to the detection CNN), as in similar works [43], [44]. Establishing an acceptable threshold is challenging, as it may vary depending on the clinical context, such as the amount of additional time readers are willing to dedicate to discarding false-positive predictions.

The error analyses that were conducted in sections 3.1 and 3.2 uncovered some interesting patterns: first, as similarly shown in other studies [45], the majority of missed aneurysms were small. Specifically, 93% of all missed UIAs were smaller than 3.5 mm for the junior (when averaging across AI-Assisted and Unassisted setting), and 91% of all missed UIAs were smaller than 3.1 mm for the senior. Second, we witnessed a strong "satisfaction of search" effect. Although this is a known issue in radiology [10], to the best of our knowledge this had never been shown for aneurysm detection in TOF-MRA. Third, the FP predictions of the junior radiologist were predominantly located in the ICA and ACA/Pcom/Posterior regions (84% of FPs), rather than in the MCA (16%), though this finding could be reader-dependent and not generalizable.

The analysis of the confidence scores highlighted two trends: on one hand, we found no significant difference between the two settings, indicating that the AI assistance does not influence the readers' confidence; on the other hand, the confidence of the junior was lower for the FPs than for the TPs, indicating that the reported False Positives were doubtful cases which would have been anyway re-checked by a senior during clinical practice.

Overall, our findings differ from those reported by the similar work of Park et al. [46], who observed a significant improvement in physician performance with AI assistance. Several key methodological differences may account for this discrepancy. First, Park et al. restricted their analysis to aneurysms larger than 3 mm, which are known to be more easily detectable by both clinicians and algorithms. In contrast, our test set imposed no size restriction, presenting a more challenging scenario; 22 aneurysms were smaller than 3 mm. Second, the diagnostic task in Park et al. was limited to providing a binary subject-level label (i.e., presence or absence of at least one aneurysm), while in our study, radiologists were asked to localize and report each aneurysm individually. This more granular task is susceptible to effects such as satisfaction of search, as discussed above. Finally, differences in imaging modality (CTA vs. MRA), network architecture, and reader cohort may have also contributed to the observed variation in results.

This study has several limitations. First, the ratio of patients with UIAs in our test set (37%) is higher than the one encountered during routine clinical practice. Second, the prevalence of patients with UIAs is different between our training and test set; further balancing this ratio (e.g., by adding more controls in the training set) might change final detection results. Nonetheless, our sample size remains substantially higher than most related studies. Third, both readers of this study were specialized in neurovascular applications, which could have potentially biased results by setting the sensitivity bar high. Fourth, the computation of the *overlay* image was performed offline (not right after the MR acquisition) and then uploaded to the original patient folders in the PACS system. This solution is impractical and should be replaced by an inline solution that is triggered right after the acquisition and directly loads the sequence in the patient folder. Fifth, despite the multi-center transfer learning approach adopted with the ADAM dataset, our test set is composed of subjects belonging to a single hospital (Lausanne University Hospital) and a single vendor (Siemens Healthineers, Forchheim, Germany). Generalization to external sites with different patient populations and scanners remains to be evaluated. Sixth, even though our 3D-Unet was optimized for aneurysm detection by adding prior anatomical information, using more complex model architectures (e.g., ResNet, Transformers) or different modalities [47], [48], [49] might have led to higher detection performances. Similarly, a larger training set is likely to yield improved detection, and therefore change the results of this study.

Considering the above-mentioned results and limitations, future works should focus on increasing sensitivity up to a level which is clinically helpful, while further reducing the false positive rate in order to reduce reading time. In addition, assessing the change in performances for a third general radiologist with no expertise in neurovascular applications would enrich the analysis. Furthermore, future works should include larger, more diverse, and multi-vendor patient populations.

In conclusion, while our AI-based CAD system demonstrated strong performance on the test set, its integration into a clinical setting did not lead to significant improvements in diagnostic sensitivity for either the junior or senior radiologist. Additionally, the AI assistance increased reading time without affecting the confidence of the readers. These findings highlight the need for thorough clinical testing to assess the real-world impact and workflow integration of AI algorithms in radiology. We believe this work could serve as reminder for the scientific community to emphasize the importance of clinical validation, ensuring that AI tools not only excel in controlled environments but also offer tangible benefits in clinical practice.

# Declaration of Generative AI and AI-assisted technologies

During the preparation of this work, the author(s) used **perplexity.ai** and **ChatGPT** to rephrase some paragraphs and improve readability. After using this tool, the author(s) reviewed and edited the content as needed and take(s) full responsibility for the content of the published article.

# Acknowledgements

This work is supported by interdisciplinary fund of the Faculty of Biology and Medicine of the Lausanne University, the CIBM Center for Biomedical Imaging, and the Swiss National Science Foundation (grant numbers: 185872 for OE).